\documentstyle[psfig,preprint,aps]{revtex}

\newcommand{\sla}{\!\!\!\!/ \,}

\def\beq{\begin{equation}}
\def\eeq{\end{equation}}
\def\bea{\begin{eqnarray}}
\def\eea{\end{eqnarray}}

\begin{document}
\hoffset-1cm
\draft

\title{Quark Propagation in a Quark-Gluon Plasma with Gluon Condensate
\footnote{Supported by DFG}}

\author{Andreas Sch\"afer and Markus H. Thoma\footnote{Permanent address:
Institut f\"ur Theoretische Physik, Universit\"at Giessen, 35392 Giessen, 
Germany}}
\address{Institut f\"ur Theoretische Physik, Universit\"at Regensburg,
93040 Regensburg, Germany}

\date{\today}

\maketitle

\begin{abstract}

We present a calculation of the thermal quark propagator taking the gluon 
condensate above the critical temperature into account. The quark dispersion
relation following from this propagator, describing two 
massive modes, is discussed.
\end{abstract} 

\bigskip

{\hspace*{1.1cm}Keywords: Quark-gluon plasma, gluon condensate, quark 
dispersion relation}

\narrowtext
\newpage
Finite temperature QCD is used to describe the properties of a quark-gluon 
plasma (QGP), which is supposed to exist in the early stage of the Universe 
and in the fireball of a relativistic heavy ion collision.
Besides lattice QCD also perturbation theory has been applied to determine
phenomenologically relevant properties of this state. In
contrast to lattice calculations the latter method is able to deal also with
dynamical properties, a finite baryon density and non-equilibrium situations.
As an important example the dispersion relations for quarks and gluons
in a quark-gluon plasma have been determined in this way\cite{ref1,ref2,ref3}.
However, using only bare propagators gauge dependent and infrared divergent 
results have been found in many cases. In order to avoid these problems 
(at least partially), Hard Thermal Loop (HTL) resummed propagators and 
vertices have to be used within an effective perturbation theory 
\cite{ref4,ref5}. However, these perturbative methods could be reliable
at best for temperatures far above the critical temperature, where the 
temperature dependent running coupling constant becomes small. 

In the present letter we investigate the consequences of a non-perturbative
quantity, namely the gluon condensate measured in lattice QCD above
the critical temperature $T_c$ \cite{ref5a}, on the quark propagator and 
the quark dispersion relation. 
Our approach is similar to the QCD sumrule approach to hadron physics at zero
temperature. We will combine purely non-perturbative input from lattice QCD
as parametrized by temperature-dependent condensates and purely perturbative
results as obtained e.g. in HTL calculations. To do this consistently, i.e. to 
avoid double counting, we have to subtract all perturbative parts in our
calculation involving the condensates.

The gluon condensate above $T_c$ is given by 
the difference of the zero temperature condensate and the interaction 
measure $\Delta =(\epsilon -3p)/T^4$, where $\epsilon$ is the 
energy density and $p$ the pressure of the QGP \cite{ref6,ref7},
\beq
\langle G^2\rangle _T = \langle G^2 \rangle _{T=0} - \Delta T^4,
\label{e1}
\eeq
where $\langle G^2\rangle _T = (11\alpha _s)/(8\pi )\, \langle 
{G^a_{\mu \nu}}^2 \rangle _T $ \cite{ref6} and $\langle G^2\rangle _{T=0}
= (2.5\pm 1.0)\, T_c^4$ \cite{ref7}.

To lowest order the interaction of a quark with the gluon condensate
is given by the self energy diagram shown in Fig.1. For massless
bare quarks it reads at finite temperature using the imaginary time 
formalism and the notation $P\equiv (p_0,{\bf p}),\; p\equiv |{\bf p}|$ 
\beq
\Sigma (P)=\frac{4}{3}\, i\, g^2\> iT\sum_{k_0=2\pi inT}\int 
\frac{d^3k}{(2\pi)^3} \tilde D_{\mu \nu}(K)\, \gamma ^\nu
\, \frac{P\sla -K\sla}{(P-K)^2}\, \gamma ^\mu,
\label{e2}
\eeq
where $\tilde D_{\mu \nu}=D^{full}_{\mu \nu}-D^{pert}_{\mu \nu}$ is the 
non-perturbative gluon propagator containing the gluon condensate. 
The perturbative gluon propagator has been subtracted since we are not 
interested in the contribution to the quark self energy coming from the bare 
gluon propagator, which is not related to the gluon condensate
but is contained in the HTL part. The diagram 
of Fig.1 together with the subtracted gluon propagator $\tilde D_{\mu \nu}$
has been used already in a zero temperature calculation \cite{ref9}.
Owing to the subtraction of the perturbative propagator, $\tilde D_{\mu \nu}$
is gauge independent. 

The most general expression for the fermion self energy (in the case of
a vanishing bare fermion mass) in the rest frame of the heat bath
is given by \cite{ref3}
\beq  
\Sigma (P)=-a(p_0,p)P\sla -b(p_0,p)\gamma _0
\label{e3}
\eeq
with the scalar functions
\bea
a & = & \frac{1}{4p^2}\> \left [tr(P\sla \Sigma )-p_0\, tr(\gamma _0\Sigma)
\right ],\nonumber \\
b & = & \frac{1}{4p^2}\> \left [P^2\, tr(\gamma _0\Sigma )-p_0\, 
tr(P\sla\Sigma)\right ].
\label{e4}
\eea

The most general ansatz for the non-perturbative gluon propagator,
reads \cite{ref8}
\beq
\tilde D_{\mu \nu} (K)=\tilde D_L(k_0,k)\, P^L_{\mu \nu} + \tilde D_T (k_0,k)\,
P^T_{\mu \nu},
\label{e5}
\eeq
where the longitudinal and transverse projectors are given by
\bea
P^L_{\mu \nu} & = & \frac{K_\mu K_\nu}{K^2}-g_{\mu \nu}-P^T_{\mu \nu}, 
\nonumber \\
P^T_{\mu 0} & = & 0, \; \; \; \; \; P^T_{ij}=\delta _{ij}-\frac{k_ik_j}{k^2}.
\label{e6}
\eea

In order to relate the propagator (\ref{e5}) to the gluon condensate
we follow the zero temperature calculation \cite{ref9} and expand the 
quark propagator in (\ref{e2}) for small loop momenta $K$.
Keeping only terms which are bilinear in $K$, we can relate the gluon 
condensate to moments of the gluon propagator \cite{ref9}. This approach,
known as plane wave method, has been widely used in QCD sum rules 
calculations \cite{ref9a}. At zero temperature it reproduces the well known 
results for the quark self energy as discussed in \cite{ref9}. Here we 
assume that it can also be applied to the finite temperature case.

At finite temperature, where $k_0$ and $k$ have to be treated separately,
new contributions containing bilinear combinations of $k_0$ and $k$ 
appear, which cannot be related to the gluon condensate. However,
since $|k_0|\ll p$ as a consequence of the plane wave method and
since $T>T_c$, it should be a good approximation 
to limit ourselves to the lowest Matsubara mode 
$k_0= 2\pi inT=0$ as long as $p$ is not much larger than $T_c$. This 
approximation is a crucial step as it is necessary to relate at finite
temperature the non-perturbative gluon propagator (\ref{e5}) to the 
condensates (\ref{e8}). 

Then expanding the quark propagator for small three momenta $k\ll p$ leads to 
\bea
a & = & -\frac{4}{3}\, g^2\, \frac{1}{P^6}\, T\, \int \frac{d^3k}{(2\pi)^3}  
\Biggl [\left (\frac{1}{3}p^2-\frac{5}{3}p_0^2\right )\, k^2\, \tilde D_L(0,k)
+\left (\frac{2}{5}p^2-2p_0^2\right )\, k^2\, \tilde D_T(0,k) \Biggr ]
\nonumber \\
b & = & -\frac{4}{3}\, g^2\, \frac{p_0}{P^6}\, T\, \int \frac{d^3k}{(2\pi)^3}  
\Biggl [\frac{8}{3}p_0^2\, k^2\, \tilde D_L(0,k)-\frac{16}{15}p^2\, k^2\, 
\tilde D_T(0,k) \Biggr ].
\label{e7}
\eea

The moments of the longitudinal and transverse gluon propagator in (\ref{e7})
can be expressed by the chromoelectric and chromomagnetic condensates
similarly as in \cite{ref9}:
\bea
\langle {\bf E}^2\rangle _T & = &  8T\> \int \frac{d^3k}{(2\pi)^3}\>  k^2\, 
\tilde D_L(0,k)+O(g), 
\nonumber \\
\langle {\bf B}^2\rangle _T & = & -16T\> \int \frac{d^3k}{(2\pi)^3}\>  k^2\, 
\tilde D_T(0,k)+O(g). 
\label{e8}
\eea

Keeping terms proportional to $k_0^2$ in (\ref{e7}), taking the zero 
temperature limit, and replacing $iT\int d^3k/(2\pi)^3 \rightarrow \int 
d^4K/(2\pi )^4$ and $\tilde D_L(k_0,k)=\tilde D_T(k_0,k)=-\tilde D(K^2)$,  
(\ref{e7}) and (\ref{e8}) reproduce the zero temperature results \cite{ref9},
$a=-g^2 \langle {G^a_{\mu \nu}}^2 \rangle _{T=0}/(36P^4)$ and $b=0$.  

The condensates defined in Minkowski space are related to the space and 
timelike plaquette expectation values, $\Delta _\sigma$ and $\Delta _\tau$, 
measured on a lattice in the case of a pure SU(3) gauge theory \cite{ref7}, by
\bea
\frac {\alpha _s}{\pi}\, \langle {\bf E}^2\rangle _T & = & \frac{4}{11}\, 
T^4\, \Delta _\tau - \frac{2}{11}\, \langle G^2 \rangle _{T=0}, \nonumber \\
\frac {\alpha _s}{\pi}\, \langle {\bf B}^2\rangle _T & = & -\frac{4}{11}
\, T^4\, \Delta _\sigma +\frac{2}{11} \, \langle G^2 \rangle _{T=0}.
\label{e9}
\eea 
The plaquette expectation values are related to the interaction measure
$\Delta $ of (\ref{e1}) by $\Delta =\Delta _\sigma +\Delta _\tau$. 

Using the results found for $\Delta _{\sigma ,\tau}$ in Ref.\cite{ref7}
we find that the electric condensate $(\alpha _s/\pi)\, \langle {\bf E}^2
\rangle _T$ increases like about $T^{3.5}$
between $T=1.1\, T_c$ and $4T_c$, whereas the magnetic condensate 
$(\alpha _s/\pi)\, \langle {\bf B}^2\rangle _T$ is close to zero
up to $T=2\, T_c$ and increases strongly afterwards. These results differ
from previous lattice calculations, where approximately temperature 
independent electric and magnetic condensates of equal size 
have been found above and close to $T_c$\cite{ref10}. 

In the perturbative regime ($g\rightarrow 0$) $\Delta _\tau =-\Delta _\sigma
=11g^2/30+O(g^4)$ holds \cite{ref7}. This result corresponds to the 
Stefan-Boltzmann limit: $\epsilon _{SB}=(\langle {\bf E}^2\rangle _{SB}
+\langle {\bf B}^2\rangle _{SB})/2=(2\pi /11\alpha _s)\, (\Delta _\tau
-\Delta _\sigma )\, T^4=(8\pi ^2/15)\, T^4$. In order to obtain the condensate
in the perturbative regime, the Stefan-Boltzmann contribution
has to be subtracted from (\ref{e9}) in the limit $g\rightarrow 0$ 
\cite{ref10}. Then the condensates are of order $g^4T^4$, describing 
perturbative corrections to the ideal gas limit. It should be noted
that the subtraction of
the Stefan-Boltzmann contribution, defined in the limit $g\rightarrow 0$,
does not change the results for the condensates near $T_c$ since there 
the plaquette expectation values are non-perturbative and cannot be written
as being proportional to $g^2$ plus higher orders. Subtracting the
Stefan-Boltzmann contribution of $\Delta _{\sigma,\tau}$ at a finite
value of $g$ would lead to an explicit $g$-dependence of the condensates 
in (\ref{e9}). 

The quark propagator containing the gluon condensate follows from the 
self energy as $S(P)=1/(P\sla -\Sigma )$. The analogous calculation
for $T=0$, which was performed in \cite{ref9}, leads to a gauge dependent
quark propagator. The gauge dependent term was found to be proportional
to the quark condensate. As the latter vanishes in our case ($T>T_c$)
due to chiral symmetry restoration we end up with a gauge independent result.  

Decomposing this propagator according to
its helicity eigenstates it can be written as \cite{ref11}
\beq
S(P)=\frac{\gamma _0-\hat p\cdot {\bf \gamma}}{2D_+(P)}+\frac{\gamma _0+
\hat p\cdot {\bf \gamma}}{2D_-(P)},
\label{e10}
\eeq
where
\beq
D_\pm(P)=(-p_0\pm p)\, (1+a) - b.
\label{e11}
\eeq

The dispersion relation of a quark interacting with the thermal gluon
condensate in a QGP is given by the roots of $D_\pm (P)$. Combining
(\ref{e7}), (\ref{e8}), and (\ref{e9}) the dispersion relations
shown in Fig.2 for $T=1.1\, T_c$, for $T=2\, T_c$, and  
for $T=4\, T_c$ are found. It should be noted that these dispersion 
relations follow from lattice results in the case of a pure gauge
theory. Unfortunately lattice calculations with dynamical quarks for the 
thermal gluon condensate are not reliable so far \cite{ref12}. However,
our results for the dispersion relations are insensitive to small
variations in the values of the condensates.

As shown in Fig.2 there are two real positive solutions
of $D_\pm(P)=0$, where the upper curve $p_0^+$ corresponds to
solutions of $D_+(P)=0$ and the lower curve $p_0^-$ to $D_-(P)=0$. Similar
as in the case of the dispersion relation following from the HTL resummed
quark propagator \cite{ref11} the dispersion relation $p_0^-$ describes 
the propagation of a quark mode with a negative helicity to chirality ratio 
(plasmino) which is absent in the vacuum. As in the HTL case the plasmino 
branch shows a minimum leading to Van Hove singularities in the 
soft dilepton production rate \cite{ref11}. 

The plasmino mode rapidly approaches the free dispersion for increasing
momenta, indicating that it is a purely collective long wave-length
mode as in the case of the HTL dispersion \cite{ref12a}. (The residue of 
this pole being proportional to $(p_0^2-p^2)^3$ for large momenta, vanishes 
for $p\gg T$.) The $p_0^+$-mode, on the other 
hand, is given by $p_0^+=p+c_1$ for large $p\gg T$, where 
$c_1=[(2\pi/9)\alpha_s\, (\langle {\bf E}^2\rangle _T +\langle 
{\bf B}^2\rangle _T/5)]^{1/4}$.

Both branches are situated
above the free dispersion relation $p_0=p$ and start from a common  
effective quark mass 
\beq
m_{eff}=\left [\frac{2\pi ^2}{3}\, \frac{\alpha_s}{\pi} \left
(\langle {\bf E}^2\rangle _T +\langle {\bf B}^2\rangle _T \right )
\right ]^{1/4}.
\label{e12}
\eeq
This effective mass is given by $m_{eff}=1.15\, T$ between $T=1.1\, T_c$
and $4\, T_c$. Thus it is independent of $g$. For small momenta 
$p\rightarrow 0$ the dispersion relation behaves like
$p_0^\pm=m_{eff}\pm c_2\, p$, where 
$c_2=(3/4)\,\langle {\bf E}^2\rangle _T/(\langle {\bf E}^2\rangle _T 
+\langle {\bf B}^2\rangle _T)$. 
  
For high temperatures the effective mass is proportional to $gT$
since the condensates in (\ref{e9}) are of order $g^4T^4$ once that the
Stefan-Boltzmann contribution has been subtracted. Therefore
it is of the same order as the HTL quark mass  $m_{HTL}=gT/\sqrt{6}$. 
Although this contribution to the effective quark mass is a 
perturbative correction, we do not expect it to be identical to
the HTL result. For the loop momentum of the quark self energy using
the HTL approximations is hard ($k\gg p$), whereas it is soft ($k\ll p$)
in our case. 
 
Let us note here that the HTL mass has been used in related studies 
considering quarks and gluons in the QGP as quasiparticles 
in order to explain the lattice data for the interaction measure $\Delta $ 
\cite{ref13}. 

As a possible application the quark dispersion and the effective 
quark propagator containing the
gluon condensate might be used to determine the dilepton and photon
production from a QGP. To lowest order the dilepton production rate 
is derived from the Born term (quark-antiquark annihilation),
where the quark dispersion could be used for the quark-antiquark pair. 
The dilepton production taking into account 
a finite temperature gluon condensate above $T_c$ has been discussed
recently in Ref.\cite{ref15}. In contrast to the HTL propagator
our effective propagator has no imaginary part below 
the light cone. Therefore there is
no contribution of the photon self energy to the photon production rate
at the one-loop level using an effective quark propagator as it is the
case within the HTL resummation scheme \cite{ref16}. However, the 
effective propagator could be used in the tree level calculation
of the photon rate, i.e. for the Compton scattering and quark-antiquark 
annihilation (with gluon emission) diagrams, together with the dispersion 
relation for the external quarks.   

Finally we want to comment on the possibility of deriving a gluon
propagator containing a gluon condensate at finite temperature.
This would be of great interest as the non-perturbative gluon propagator 
might show a static magnetic screening, which is not present in
perturbative calculations leading to infrared divergences even when HTL 
resummed propagators are used (see e.g. \cite{ref5}). Unfortunately, the gluon 
propagator containing a gluon condensate is a very complicated object 
already at zero temperature \cite{ref17}. In order to preserve the 
Slavnov-Taylor identities, i.e. the transversality of the gluon self energy
containing the gluon condensate, higher order condensates have to be included. 
These condensates, however, are not known at finite temperature. 
Furthermore this gluon propagator exhibits a dependence on the gauge 
parameter rendering a physical interpretation of its poles doubtful.

Summarizing, we have shown that the presence of a gluon condensate above 
$T_c$, as suggested by lattice calculations, leads to an interesting
quark dispersion relation, which exhibits a large similarity to
the dispersion resulting from the HTL quark propagator. In both
cases two massive branches are found, where the plasmino mode shows a 
minimum at a finite value of the momentum. For large momenta all branches
approach the free dispersion relation. However, our 
effective quark mass differs from the HTL mass, which is proportional to $gT$.
For temperatures close to $T_c$ it is proportional to $T$, while it is
proportional to $gT$ in the perturbative regime.

\vspace*{1cm}

\centerline{\bf ACKNOWLEDGMENTS}

We would like to thank V. Braun, M. G\"ockeler, U. Heinz, F. Karsch, 
M. Lavelle, S. Leupold, M. Maul, B. M\"uller, and M. Schaden for 
stimulating and helpful discussions.

\begin{figure}

\vspace*{3cm}

\centerline{\psfig{figure=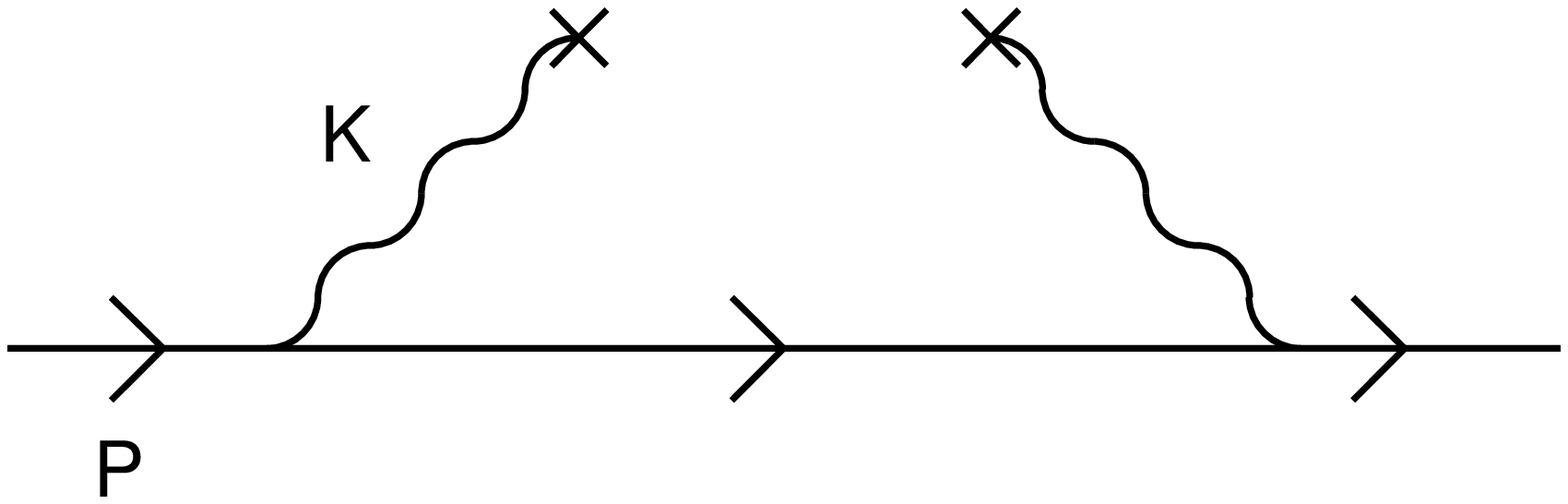,width=8cm}}
\caption{Quark self energy containing a gluon condensate.}
\end{figure}

\begin{figure}
\centerline{\psfig{figure=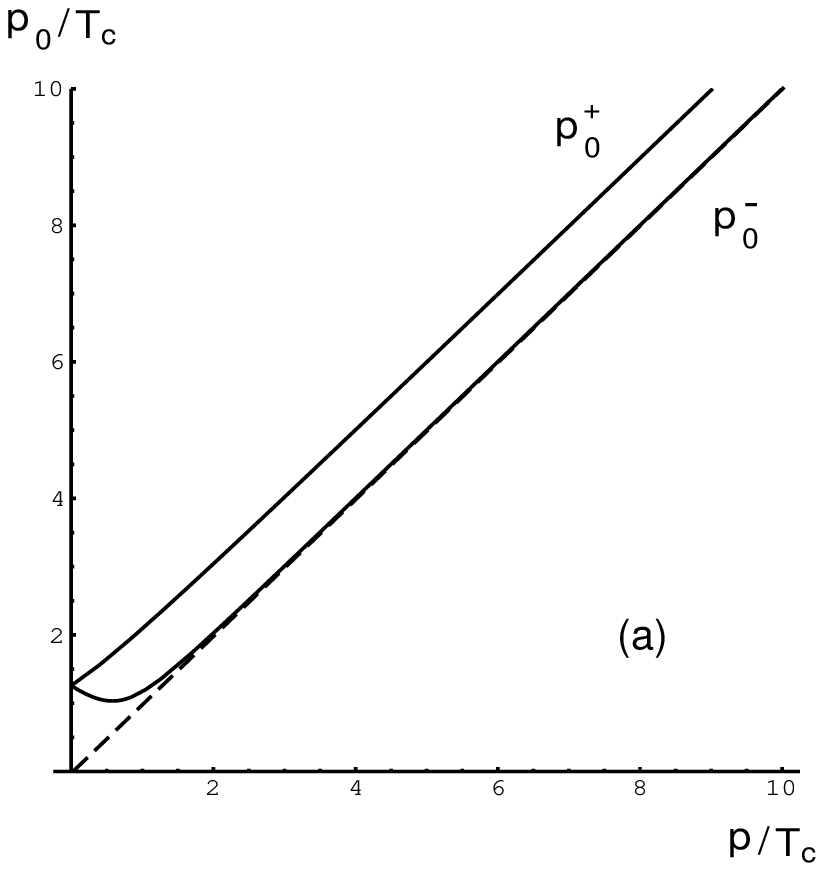,width=8cm}}
\centerline{\psfig{figure=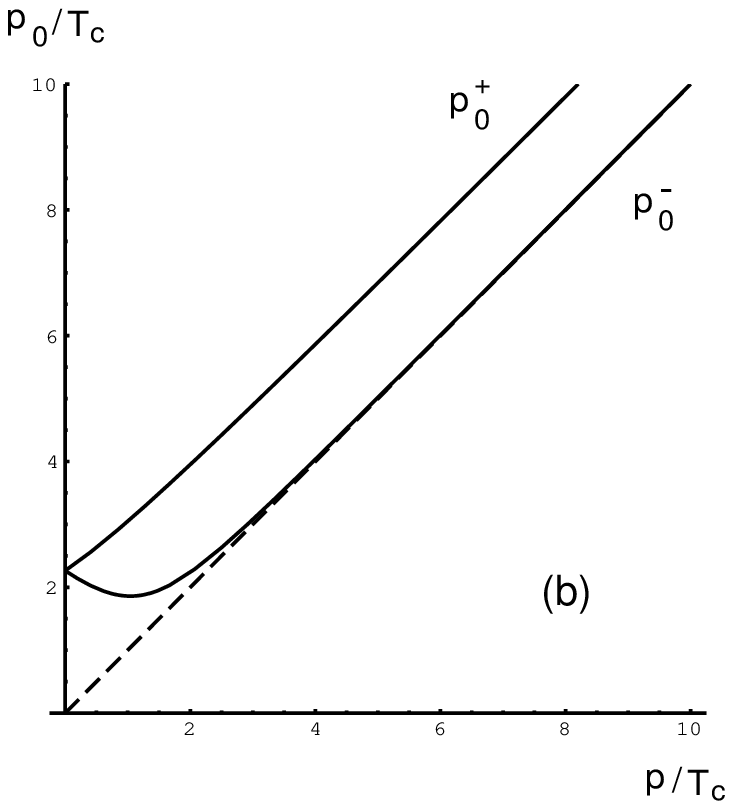,width=8cm}}
\centerline{\psfig{figure=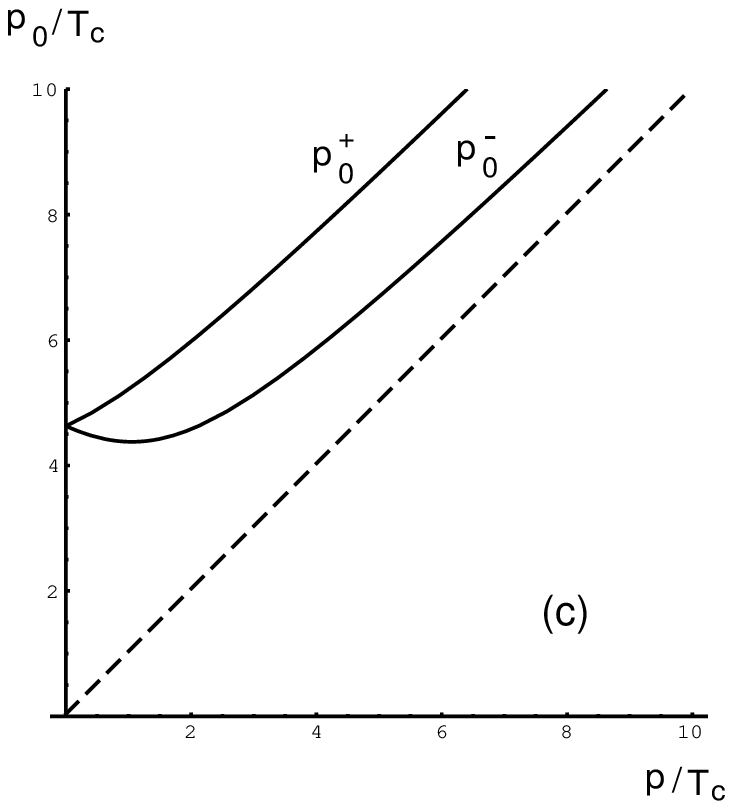,width=8cm}}
\vspace*{0.5cm}
\caption{Quark dispersion relations at $T=1.1\, T_c$ (a), $T=2 \, T_c$ (b),
$T=4\, T_c$ (c) and dispersion relation of a non-interacting massless quark
(dashed lines).}
\end{figure}

\end{document}